\documentclass[twocolumn, prl, superscriptaddress]{revtex4-1}
\usepackage{amssymb}
\usepackage{amsmath}
\usepackage{epsfig}
\usepackage{graphics, graphicx}
\usepackage{bbold}
\usepackage{psfrag}
\usepackage{mathcomp}
\usepackage{subfigure}
\usepackage{verbatim}
\usepackage{color}
\usepackage[colorlinks,citecolor=blue]{hyperref}
\def\cp#1{\mathbf{#1}}

\begin{document}
	\title{Quartet Superfluid in Two-dimensional Mass-imbalanced Fermi Mixtures}
	
	\author{Ruijin Liu}
	\affiliation{Beijing National Laboratory for Condensed Matter Physics, Institute of Physics, Chinese Academy of Sciences, Beijing, 100190, China}
	
	\author{Wei Wang}
	\affiliation{Beijing National Laboratory for Condensed Matter Physics, Institute of Physics, Chinese Academy of Sciences, Beijing, 100190, China}
	\affiliation{School of Physical Sciences, University of Chinese Academy of Sciences, Beijing 100049, China}
	
	\author{Xiaoling Cui}
	\email{xlcui@iphy.ac.cn}
	\affiliation{Beijing National Laboratory for Condensed Matter Physics, Institute of Physics, Chinese Academy of Sciences, Beijing, 100190, China}
	
	\begin{abstract}
		Quartet superfluid (QSF) is a distinct type of fermion superfluidity that exhibits high-order correlation beyond the conventional BCS pairing paradigm. 		In this Letter, we report the emergent QSF in 2D mass-imbalanced Fermi mixtures with two-body contact interactions.
		This is facilitated by the formation of quartet bound state  in vacuum that consists of a light atom and three heavy fermions.
				For an optimized heavy-light number ratio $3:1$, we identify QSF as the ground state in a considerable parameter regime of mass imbalance and 2D coupling strength. 		Its unique high-order correlation can be manifested in the momentum-space crystallization of pairing field and density distribution of heavy fermions.
		Our results can be readily detected in Fermi-Fermi mixtures nowadays realized in cold atoms laboratories, and meanwhile shed light on exotic superfluidity in a broad context of mass-imbalanced fermion mixtures.
	\end{abstract}
	
	\date{\today}
	\maketitle
A basic idea to achieve the superfluidity of fermions is to bind 	an even number of fermions into a composite boson and then let bosons condense. 	A classic example  is the Bardeen-Cooper-Schrieffer (BCS) superfluid in terms of the condensation of Cooper pairs~\cite{BCS}, which reflects the two-body correlation among spin-1/2 fermions and has achieved great success not only in solid states, but also in cold atoms~\cite{review_1,review_2} and nuclear matters~\cite{review_nuclear1,review_nuclear2}. 	Going beyond the BCS framework, a fascinating yet challenging direction is to  engineer superfluids with higher-order correlations. Along this direction, a leading case is the quartet superfluid (QSF), a distinct type of fermion superfluidity based on the condensation of four-fermion clusters.  Previous studies have revealed  QSF in spin-3/2 fermions~\cite{Wu}, nuclei with $\alpha$-particle condensation~\cite{quartet_nuclear1,quartet_nuclear2,quartet_nuclear3,quartet_nuclear4,Tajima1},  biexciton condensates~\cite{Tajima2},  and various systems hosting charge-4e superconductivity~\cite{Volovik,Babaev2004, Babaev2010, Babaev2021, Kivelson, Fu, Yao1, Yao2, charge4e_expt, Hu, Wang}. 	However, stringent conditions are required therein such as multi-components, multi-body interactions or pair fluctuations under particular symmetries, which make the experimental exploration of QSF rather rare and difficult in practice.

	Recently, mass-imbalanced Fermi mixtures realized in ultracold gases, such as $^{40}$K-$^{6}$Li~\cite{K_Li1, K_Li2, K_Li3}, $^{161}$Dy-$^{40}$K~\cite{Dy_K1, Dy_K2} 
	and $^{53}$Cr-$^{6}$Li~\cite{Cr_Li, Cr_Li2, Cr_Li3}, offer a much easier platform for achieving QSF. The predominant few-body correlation in these systems can be inferred from the formation of universal few-body clusters that consist of a light atom and several heavy fermions. Each cluster bound state requires the  heavy-light mass ratio beyond certain critical value~\cite{KM, Blume, Petrov,  Pricoupenko, Parish,Cui, KM_1D,Mehta_1D,Petrov_1D} but still small enough to avoid any Efimovian binding~\cite{Efimov,Castin,Petrov}. These clusters are therefore believed to be elastically stable under collision. Physically, their formation is due to a long-range heavy-heavy attraction mediated by the light atom, which competes with a repulsive centrifugal barrier in $p$-wave channel~\cite{KM, BOA}.
	As such, the critical mass ratio to support a tetramer bound state (a quartet) is found to be quite high in 3D~\cite{Blume, Petrov}, but is sufficiently low in 2D~\cite{Cui} so as to be accessible by all Fermi mixtures listed above. This quartet formation has been shown to fundamentally change the destiny of Fermi polaron compared to equal mass case, when increasing the attraction between light impurity and heavy majorities~\cite{Cui2}.
	Then the ultimate question is how would the quartet affect the many-body property of heavy-light mixtures? In particular, can QSF emerge as ground state? If so, this will be the {\it simplest} fermion system so far to support QSF, i.e., with only two components and under two-body contact interactions. Such a system would be much more convenient to manipulate experimentally.
	
	\begin{figure}[t]
		\includegraphics[width=8.5cm]{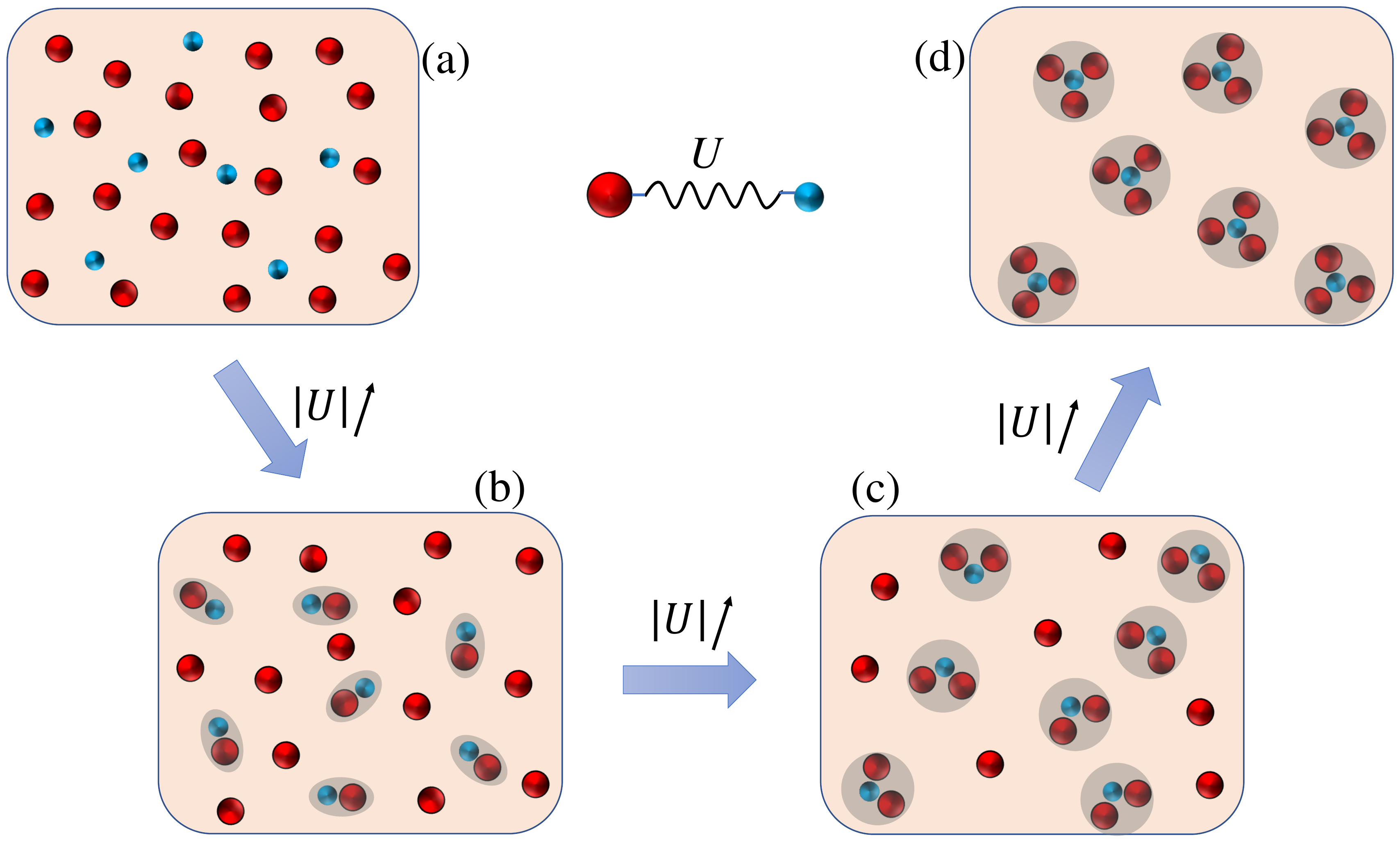}
		\caption{(Color Online). Schematics for different phases in heavy(red)-light(blue) fermion mixtures with number ratio $3:1$. As increasing the attraction strength $|U|$, the system undergoes a sequence of phase transitions from the normal mixture (a) to mixed pairing superfluid and normal state (b), mixed trimer liquid and normal state (c) and finally to quartet superfluid (d).  }  \label{fig_schematic}
	\end{figure}
	
\begin{figure}[t]
		\includegraphics[width=8.5cm]{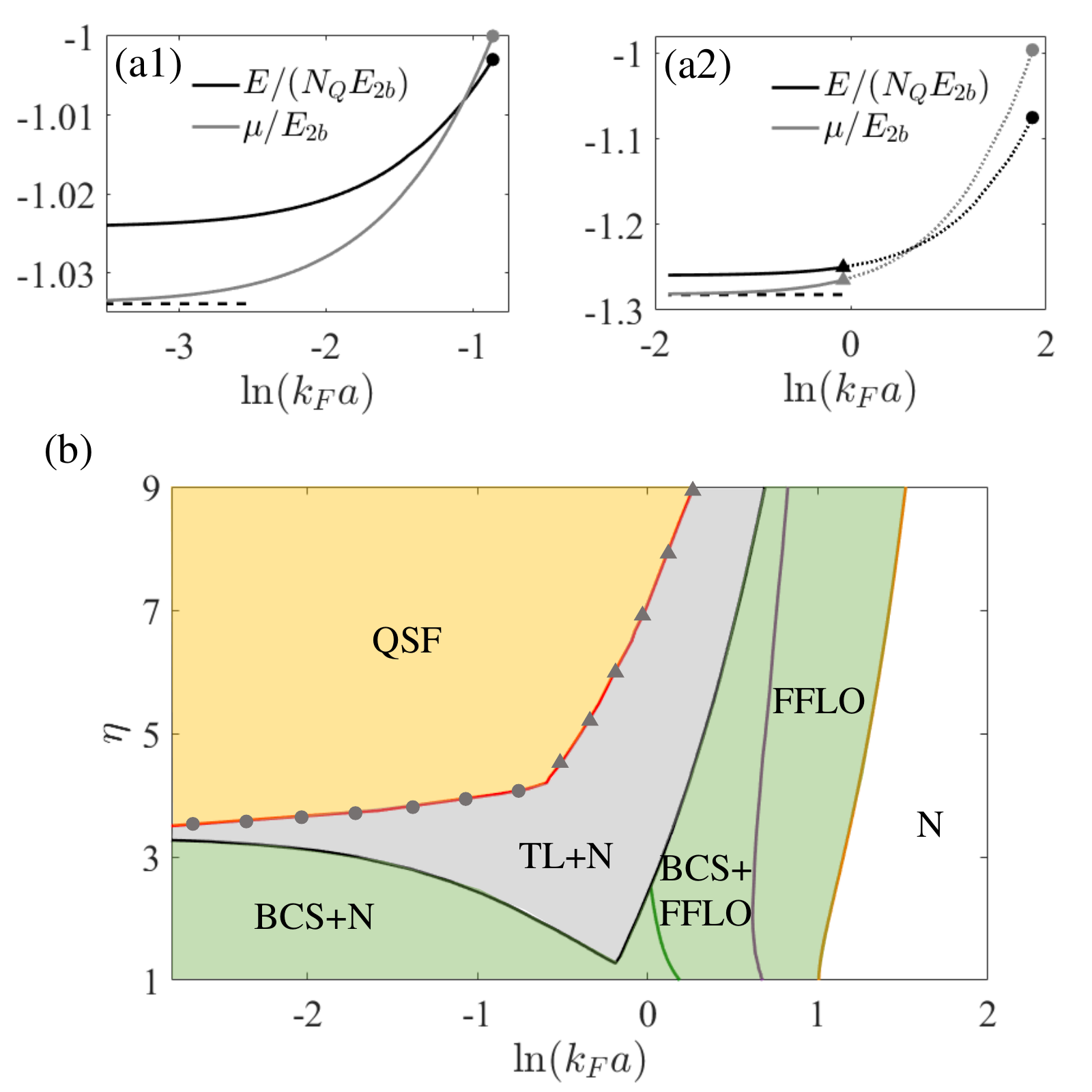}
		\caption{(Color Online). Emergence of quartet superfluid in 2D heavy-light fermion mixtures with number ratio $3:1$. (a1,a2) Energy per quartet $E/N_Q$ and chemical potential $\mu$ of the quartet superfluid as functions of coupling strength $\ln(k_Fa)$ for mass ratio $\eta=161/40$ (a1) and $40/6$ (a2). Here the energy unit is the two-body binding energy $E_{2b}$, and the horizontal dashed line shows the quartet binding energy in vacuum~\cite{Cui}. The circles denote the termination of self-consistent solutions from Eqs (\ref{gap_eq},~\ref{N_eq}). The triangles in (a2) mark the location when the Pauli principle starts to be violated, and thus the solutions from triangles to circles are unphysical. (b) Phase diagram in  ($\ln(k_Fa),\ \eta$) parameter plane. The phases (from left to right) are the quartet superfluid ('QSF', yellow area), mixed trimer liquid and normal state ('TL+N', gray), states involving pairing superfluids ('BCS+N', 'BCS+FFLO', 'FFLO', green area), and the normal state ('N', white). The QSF boundaries   with circles or triangles are in accordance with notations in (a1,a2).  		}  \label{fig_phase}
	\end{figure}

\begin{figure}[t]
		\includegraphics[width=8.5cm]{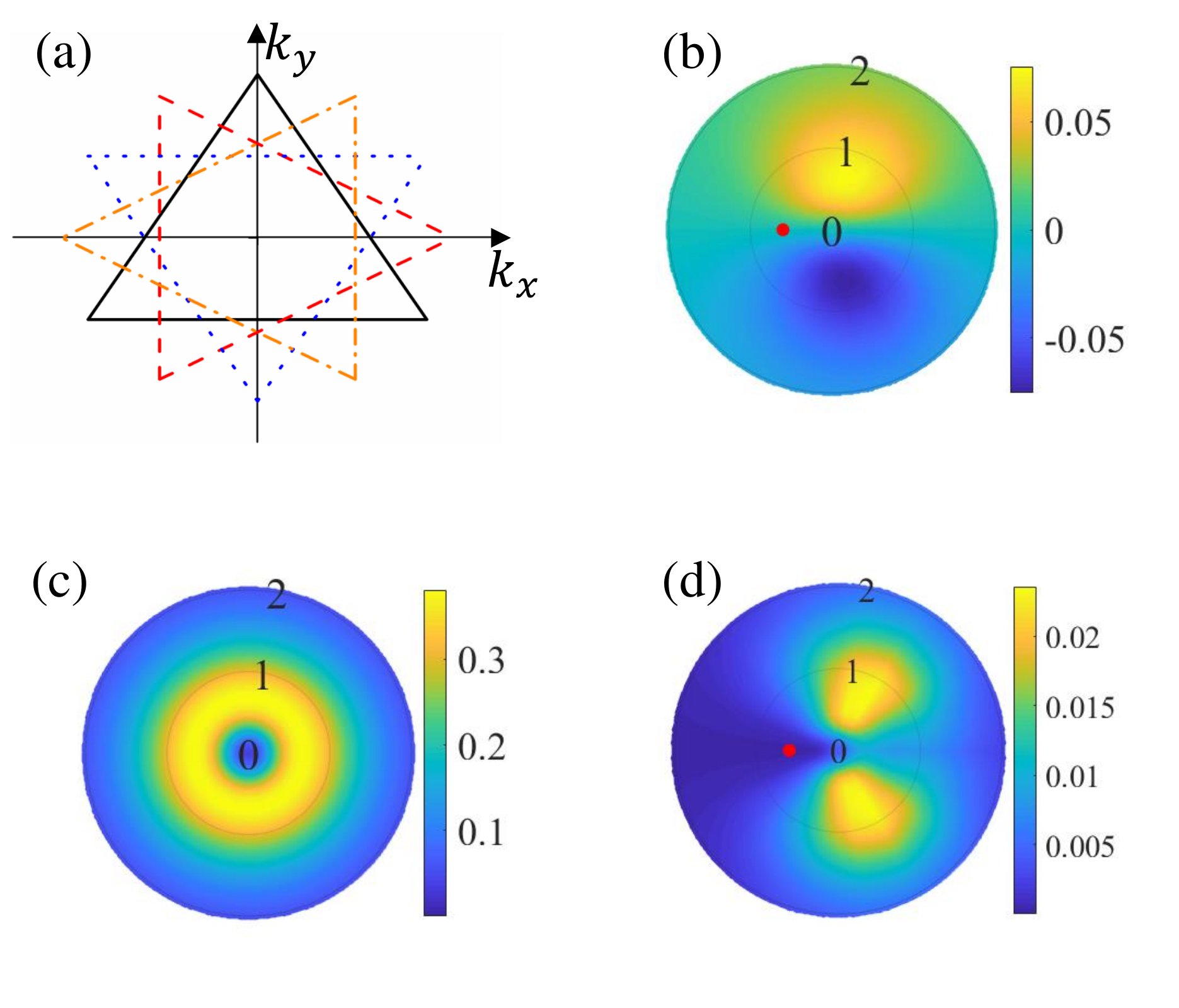}
		\caption{(Color Online). Momentum-space correlation of quartet superfluid. (a) The largest $|\psi_{{\cp k}_1{\cp k}_2{\cp k}_3}|$ with $({\cp k}_1,\ {\cp k}_2,\ {\cp k}_3)$ forming a regular triangle. Several degenerate triangles are shown, which have the same value of $|\psi|$. (b) $\Delta_{{\cp k}_0 {\cp k}}$ (in unit of $E_{2b}$) in ${\cp k}$-space, with ${\cp k}_0$ pinned at one momentum for the largest $|\psi|$ (red point). (c) Mean density distribution of heavy fermions, $\langle n_h({\cp k})\rangle$ (in unit of $S^{-1}$). (d) Density-density correlation of heavy fermions, $D_h({\cp k}_0,{\cp k})$ (in unit of $S^{-2}$), with ${\cp k}_0$  pinned  at one maximum of $\langle n_h\rangle$ (red point). In (b-d), we take $\eta=40/6$ and $\ln(k_Fa)=-0.58$.
		}  \label{fig_correlation}
	\end{figure}

	In this Letter, we unveil the emergent QSF  in 2D mass-imbalanced Fermi mixtures. 
	A variational ansatz is constructed to describe QSF with optimal  heavy-light number ratio $3:1$, which well incorporates the essential four-body correlations in a many-body setting. In contrast to previous studies of mass-imbalanced fermions that focused on pairing superfluids~\cite{Liu1, Liu3, Duan, Yip, Simons, Stoof, Calson, Conduit}, our work identifies QSF as  the most energetically favorable (ground) state under sufficient mass imbalance and attraction strength between heavy-light fermions. This suggests a sequence of phase transitions, namely, from a normal Fermi mixture to eventually a QSF, when increasing the heavy-light attractions (see Fig.~\ref{fig_schematic}).
	We have mapped out a phase diagram for QSF and other competing states as tuning the 2D coupling strength and mass imbalance (Fig.~\ref{fig_phase}). 
	Furthermore, we show that the unique high-order correlation of QSF manifests itself  in the momentum-space crystallization of pairing field and density distribution of heavy fermions (Fig.~\ref{fig_correlation}).
			Such QSF, emerging in the seemingly simple framework of mass-imbalanced two-component fermions, represents a qualitatively new kind of high-order superfluidity in strongly correlated fermionic matter. 
			Our results  can be readily probed in a number of Fermi mixtures realized in ultracold atoms, and meanwhile shed light on exotic superfluidity in a broad context of fermion systems with mass imbalance, such as the semiconducting transition metal dichalcogenides.

We start from the following Hamiltonian ($\hbar=1$):
	\begin{equation}
		H=\sum_{{\cp k}} \left(\epsilon^l_{{\cp k}} l_{{\cp k}}^{\dagger} l_{{\cp k}} + \epsilon^h_{{\cp k}} h_{{\cp k}}^{\dagger} h_{{\cp k}}\right) +\frac{g}{S} \sum_{\mathbf{q}, {\cp k}, {\cp k}^{\prime}} l_{\mathbf{q}-{\cp k}}^{\dagger} h_{{\cp k}}^{\dagger} h_{{\cp k}^{\prime}} l_{\mathbf{q}-{\cp k}^{\prime}}. \label{eq:H}
	\end{equation}
	Here $h_{{\cp k}}^{\dagger}$ and $l_{{\cp k}}^{\dagger}$ respectively create a heavy  and a light fermion  at momentum ${\cp k}$ with energy $\epsilon^{h,l}_{{\cp k}}={\cp k}^2/(2m_{h,l})$, and their mass ratio is $\eta\equiv m_h/m_l\  (>1)$; the 2D bare coupling $g$ is renormalized through $1/g=-1/S \sum_{{\cp k}}1/(\epsilon^l_{{\cp k}}+\epsilon^h_{{\cp k}}+E_{2b})$, where $S$ is the system area and $E_{2b}=(2m_r a^2)^{-1}$ is the two-body binding energy given by scattering length $a$ and reduced mass $m_r=m_lm_h/(m_l+m_h)$. In this work, we consider the most favorable heavy-light number ratio for QSF, i.e., $N_h:N_l=3:1$. Accordingly we introduce a momentum unit as $k_F=\sqrt{4\pi N_Q/S}$, with $N_Q=N_l=N_h/3$ the number of quartets.

	For a microscopic description of QSF, we utilize the exact wavefunction of a zero-momentum quartet (tetramer bound state) in vacuum that is created by
	\begin{equation}
		Q^{\dag} = \sum_{{\cp k}_1{\cp k}_2{\cp k}_3} \phi_{{\cp k}_1{\cp k}_2{\cp k}_3} l^{\dag}_{-{\cp k}_1-{\cp k}_2-{\cp k}_3} h^{\dag}_{{\cp k}_1}h^{\dag}_{{\cp k}_2}h^{\dag}_{{\cp k}_3}. \label{wf}
	\end{equation}
	Treating the quartet as a composite boson and recalling the intrinsic relation between fermion superfluidity and Bose condensation, we write down a quartet-condensed coherent state,  $e^{\lambda Q^{\dag}}$, to describe the QSF state of fermions, which can be further simplified as
	\begin{equation}
		\Psi_{\rm QSF} = \prod_{\langle{\cp k}_1{\cp k}_2{\cp k}_3\rangle} (1+\psi_{{\cp k}_1{\cp k}_2{\cp k}_3} l^{\dag}_{-{\cp k}_1-{\cp k}_2-{\cp k}_3} h^{\dag}_{{\cp k}_1}h^{\dag}_{{\cp k}_2}h^{\dag}_{{\cp k}_3})|0\rangle. \label{wf_QSF}
	\end{equation}
	Here the variational coefficients $\psi_{{\cp k}_1{\cp k}_2{\cp k}_3}$ are anti-symmetric with respect to the exchange  ${\cp k}_i\leftrightarrow {\cp k}_j$ ($i\neq j$), and  $\langle\rangle$  denotes that we avoid any double counting of ${\cp k}$-triples $\{{\cp k}_1{\cp k}_2{\cp k}_3\}$.
	
	Different from the well-known BCS ansatz ($\sim \prod_{\cp k}(1+\psi_{\cp k}l^{\dag}_{-{\cp k}}h^{\dag}_{\cp k})$) where only one ${\cp k}$-index is used to characterize each Cooper pair,  the QSF wavefunction (\ref{wf_QSF}) displays a much higher degree of freedom given three momenta (i.e., a $\cp k$-triple) in each bracket to label a quartet. This implies a higher degree of complexity in treating the many-body problem, especially when considering the Pauli effect and the non-uniqueness of fermion occupation from multi-brackets. In this Letter, as the first attempt to include quartet correlation in tackling the fermion superfluid problem,
	we shall neglect the contribution from these complexities. It can be proved that their induced corrections are of the order $\sim \psi^2$, which can be well controlled as long as all $\psi\ll 1$ (valid especially in strong coupling regime)~\cite{supple}. 	

	Under above treatment, we can expand the thermodynamic potential $\Omega\equiv \langle H-\mu N_Q \rangle_{\rm QSF}$ as a function of $\psi$, with $\mu$ introduced as the quartet chemical potential:
	\begin{eqnarray}
		\Omega&=&\sum_{\langle{\cp k}_1{\cp k}_2{\cp k}_3\rangle} \frac{|\psi_{{\cp k}_1{\cp k}_2{\cp k}_3}|^2}{1+|\psi_{{\cp k}_1{\cp k}_2{\cp k}_3}|^2} E_{{\cp k}_1{\cp k}_2{\cp k}_3 }  +\nonumber \\
		&&\sum_{\langle{\cp k}_1{\cp k}_2{\cp k}_3\rangle} \frac{\psi^*_{{\cp k}_1{\cp k}_2{\cp k}_3}}{1+|\psi_{{\cp k}_1{\cp k}_2{\cp k}_3}|^2} (\Delta_{{\cp k}_2{\cp k}_3}-\Delta_{{\cp k}_1{\cp k}_3}+\Delta_{{\cp k}_1{\cp k}_2}); \nonumber
	\end{eqnarray}
here $E_{{\cp k}_1{\cp k}_2{\cp k}_3 }=\epsilon^l_{-{\cp k}_1-{\cp k}_2-{\cp k}_3}+\epsilon^h_{{\cp k}_1}+\epsilon^h_{{\cp k}_2}+\epsilon^h_{{\cp k}_3}-\mu$, and the auxiliary function $\Delta_{{\cp k}_2{\cp k}_3}$  is defined as
	\begin{equation}
		\Delta_{{\cp k}_2{\cp k}_3}=\frac{g}{S} \sum_{{\cp k}_1} \frac{\psi_{{\cp k}_1{\cp k}_2{\cp k}_3}}{1+|\psi_{{\cp k}_1{\cp k}_2{\cp k}_3}|^2}. \label{gap}
	\end{equation}
Minimizing $\Omega$ via  $\partial \Omega/\partial \psi^*_{{\cp k}_1{\cp k}_2{\cp k}_3}=0$, we  obtain
	 \begin{eqnarray}
	 	&&\psi_{\mathbf{k}_1\mathbf{k}_2\mathbf{k}_3}\nonumber\\
	 	&=&\frac{E_{\mathbf{k}_1\mathbf{k}_2\mathbf{k}_3}-\sqrt{E_{\mathbf{k}_1\mathbf{k}_2\mathbf{k}_3}^2+4(\Delta_{\mathbf{k}_2\mathbf{k}_3}-\Delta_{\mathbf{k}_1
	 				\mathbf{k}_3}+\Delta_{\mathbf{k}_1\mathbf{k}_2})^2}}{2(\Delta_{\mathbf{k}_2\mathbf{k}_3}-\Delta_{\mathbf{k}_1
	 			\mathbf{k}_3}+\Delta_{\mathbf{k}_1\mathbf{k}_2})}. \nonumber\\
				\label{psi}
	 \end{eqnarray}
	Further utilizing (\ref{gap}) we arrive at the self-consistent equation for $\{\Delta_{{\cp k}{\cp k}'}\}$:
		\begin{equation}
		-\frac{S}{g}\Delta_{{\cp k}_2{\cp k}_3}
		=\sum_{{\cp k}_1} \frac{\Delta_{{\cp k}_2{\cp k}_3}-\Delta_{{\cp k}_1
				{\cp k}_3}+\Delta_{{\cp k}_1{\cp k}_2}}{\sqrt{E_{{\cp k}_1{\cp k}_2{\cp k}_3}^2+4(\Delta_{{\cp k}_2{\cp k}_3}-\Delta_{{\cp k}_1{\cp k}_3}+\Delta_{{\cp k}_1{\cp k}_2})^2}}. \label{gap_eq}
	\end{equation}	
The number equation $N_Q=-\partial \Omega/\partial \mu$ is written as
	\begin{equation}
		2N_Q\!=\!\!\! \sum_{\langle{\cp k}_1{\cp k}_2{\cp k}_3\rangle} \!\!\! \Big(1-\frac{E_{{\cp k}_1{\cp k}_2{\cp k}_3}}{\sqrt{E_{{\cp k}_1{\cp k}_2{\cp k}_3}^2+4(\Delta_{{\cp k}_2{\cp k}_3}-\Delta_{{\cp k}_1{\cp k}_3}+\Delta_{{\cp k}_1{\cp k}_2})^2}}\Big).
		\label{N_eq}
	\end{equation}
	
	Physically, $\Delta_{{\cp k}_2{\cp k}_3}$ in Eq.~(\ref{gap}) can be viewed as the pairing field of QSF, since it is obtained by contracting the internal degree of one heavy-light pair in the quartet while leaving the two additional heavy fermions free at ${\cp k}_2,{\cp k}_3$. This is dramatically different from the pairing field in usual BCS theory which is a constant rather than  ${\cp k}$-dependent. In fact, the right sides of Eqs.~(\ref{psi},\ref{gap_eq},\ref{N_eq}) suggest a superposed pairing field, $\tilde{\Delta}_{{\cp k}_1{\cp k}_2{\cp k}_3}\equiv \Delta_{{\cp k}_2{\cp k}_3}-\Delta_{{\cp k}_1{\cp k}_3}+\Delta_{{\cp k}_1{\cp k}_2}$, to uniquely identify a quartet in ${\cp k}$-space. In the strong coupling limit with deep quartet binding, we have $|\tilde{\Delta}_{{\cp k}_1{\cp k}_2{\cp k}_3}|\ll |\mu|$ and therefore Eq.~(\ref{gap_eq}) well reproduces the exact equation for a quartet bound state in vacuum~\cite{Petrov, Parish, Cui}. This guarantees the picture of quartet condensation in this limit with $\mu\rightarrow E_Q$,  where $E_Q$ is the binding energy of a vacuum quartet.

	We have numerically solved Eqs.~(\ref{gap_eq},~\ref{N_eq}) to obtain $\mu$ and $\{\Delta_{{\cp k}{\cp k}'}\}$ for given interaction strength $\ln(k_Fa)$ and mass ratio $\eta$. The total energy $E\equiv \langle H\rangle_{\rm QSF}$ can then be computed straightforwardly. Note that  in Eq.~(\ref{N_eq}), the summation on ${\cp k}$-triples  brings another relevant parameter $S/a^2$ to the problem, which is taken as $100$ throughout the paper. We have checked that different $S/a^2$ will not qualitatively change our results~\cite{supple}.

	In Fig.~\ref{fig_phase}(a1,a2), we take the  Dy-K and K-Li mixtures as two experimentally relevant examples and plot their corresponding energy per quartet ($E/N_Q$) and chemical potential ($\mu$) as functions of $\ln(k_Fa)$. As expected, in the strong coupling limit $\ln(k_Fa)\rightarrow -\infty$, both $E/N_Q$ and $\mu$ approach $E_Q$ (dashed horizontal lines). As moving to weaker coupling, both quantities increase up to a critical coupling strength where $\mu\sim -E_{2b}$, see circles in Fig.~\ref{fig_phase}(a1,a2), beyond which Eqs.~(\ref{gap_eq},~\ref{N_eq}) fail to produce a convergent solution. Before reaching this point, however, the solutions can become unphysical due to the violation of Pauli principle, i.e., the ${\cp k}$-space number  of heavy fermions exceeds beyond $1$ ($N^h_{\cp k}>1$), as bounded by triangles in Fig.~\ref{fig_phase}(a2). 
	Such violation can be attributed to the neglect of Pauli effect in treating $\Psi_{\rm QSF}$~\cite{supple}, which results in a quick accumulation of $N^h_{\cp k}$ beyond $1$ as the system departs from strong coupling regime.
	In the following, we shall only take the physical solutions of QSF with all $N^h_{\cp k}<1$, i.e., for $\ln(k_Fa)$ ranging from $-\infty$ to the circles in Fig.~\ref{fig_phase}(a1) and to the triangles in Fig.~\ref{fig_phase}(a2).

	We have compared the energy of QSF with all other competing states including a normal mixture, various pairing superfluids studied for mass-imbalanced fermions~\cite{Liu1, Liu3, Duan, Yip, Simons, Stoof, Calson, Conduit}, a trimer liquid proposed in 3D~\cite{Naidon}, as well as a pentamer liquid~\cite{supple}. Finally a phase diagram is mapped out in ($\ln(k_Fa),\ \eta$) plane for a fixed number ratio $N_h/N_l=3$, see Fig.~\ref{fig_phase}(b).
	The relevant phases appearing on the diagram are QSF (yellow area), mixed trimer liquid and normal state ('TL+N', gray), states involving pairing superfluids (green) and the normal mixture ('N', white). The pairing superfluids include the Fulde-Ferrell-Larkin-Ovchinnikov superfluid ('FFLO') and two phase-separated states [between
	BCS and FFLO ('BCS+FFLO') and between BCS and normal ('BCS+N')]. For 'TL+N', we have approximated it as two homogeneous Fermi seas of trimers and of excess heavy atoms, each comprising $N_l$ particles. 
	In justifying this, we require the trimer on top of a heavy Fermi sea to be a true bound state ($E_t<0$), with energy lower than that of the corresponding atom-dimer threshold  ($E_t<E_d+E_F^h$). This leads to the phase boundaries between 'TL+N' and other pairing superfluids in Fig.~\ref{fig_phase}(b), and we have confirmed that within the gray area 'TL+N' is indeed more  energetically favorable than all other available states. Similarly, we have also considered a pentamer liquid in coexistence with a light Fermi sea, and found that such state always has a higher energy than QSF and thus is not relevant (see more details in Supplemental Material~\cite{supple}).

	Importantly, Fig.~\ref{fig_phase}(b) shows that QSF represents the ground state in a considerably broad parameter region with  $\ln(k_Fa)\lesssim0$ and $\eta>\eta_{\rm Q}\sim 3.4$ ($\eta_{\rm Q}$ is the critical mass ratio to support a 2D quartet in vacuum~\cite{Cui}). Moreover, it tells that under sufficient $\eta\ (>\eta_{\rm Q})$, the system undergoes a sequence of phase transitions as increasing the heavy-light attractions, i.e., from a normal mixture at weak coupling to states involving pairing superfluid or trimer liquid, and finally ending up at QSF at strong coupling (see also Fig.~\ref{fig_schematic}). We remark here that the occurrence of these transitions is physically robust, because QSF cannot adiabatically connect to a normal Fermi-sea as interactions are reduced. This can be clearly seen from its wavefunction (\ref{wf_QSF}), which involves a fundamental restructuring of heavy-light distributions in ${\cp k}$-space and thus cannot reproduce two uncorrelated Fermi-seas by sending $\psi$ to $\infty$. It is very different from the BCS ansatz~\cite{BCS}, 	which in weak coupling limit is just a slight modification of a normal Fermi sea. 
	Therefore the BCS ansatz can well describe a smooth BCS-BEC crossover for balanced spin-$1/2$ fermions, but here a sequence of transitions are produced for mass/spin-imbalanced systems. Such difference is intrinsically due to the high-order correlation hidden in QSF, as revealed below.

	Different from all other phases in Fig.~\ref{fig_phase}(b),  QSF exhibits unique high-order correlation in ${\cp k}$-space. Such correlation originates from the internal structure of quartet wavefunction $\psi_{{\cp k}_1{\cp k}_2{\cp k}_3}$, as shown in Fig.~\ref{fig_correlation}(a), which has the largest weight if the ${\cp k}$-triple (${\cp k}_1 {\cp k}_2 {\cp k}_3$) form a regular triangle. This is consistent with the crystalline structure of a quartet in vacuum~\cite{Cui}. Physically, the triangular structure emerges because it is highly symmetric and thus provides the largest phase space for three fermions scattering within a quartet unit. Similar triangular distribution also appears in the pairing field $\Delta_{{\cp k}_0{\cp k}}$, as shown in Fig.~\ref{fig_correlation}(b), when fixing ${\cp k}_0$ at the largest $|\psi|$ (red point). Interestingly, $\Delta_{{\cp k}_0{\cp k}}\sim {\cp k}$ exhibits a chiral distribution, i.e., its sign switches depending on whether  ${\cp k}$ moves clock- or anticlock-wise from  ${\cp k}_0$.  This can be attributed to the  anti-symmetry of $\Delta_{{\cp k}_0{\cp k}}$ with respect to ${\cp k}_0\leftrightarrow{\cp k}$, as required by its definition in Eq.~(\ref{gap}).
	
	To experimentally detect above correlation, we propose measuring the density-density correlation function of heavy fermions in ${\cp k}$-space~\cite{supple}:
	\begin{equation}
		D_h({\cp k}_0,{\cp k})\equiv  \langle n_h({\cp k}_0)n_h({\cp k})\rangle - \langle n_h({\cp k}_0)\rangle \langle n_h({\cp k})\rangle. \label{eq:dd}
	\end{equation}
	Here  $\langle n_h({\cp k})\rangle$ is the mean density distribution of heavy fermions. 
	In Fig.~\ref{fig_correlation}(c,d), we show  $\langle n_h({\cp k})\rangle$ and $D_h({\cp k}_0,{\cp k})$  for a typical QSF state of K-Li mixture. We can see that $\langle n_h({\cp k})\rangle$ is  peaked at a finite $|{\cp k}|$ and shows a dip at ${\cp k}=0$, dramatically different from the  distributions of a normal Fermi sea or a pairing superfluid. 	Fixing  ${\cp k}_0$ at the peak of $\langle n_h \rangle$,  $D_h({\cp k}_0,{\cp k})$ shows two visible peaks in ${\cp k}$-space, which form a regular triangle together with ${\cp k}_0$.
		This visualizes the unique high-order correlation in QSF that is absent in all other states. For instance, one has $D_h({\cp k}_0,{\cp k})=0$ for normal state and any type of pairing superfluids.
	In cold atoms experiment, the mean density and density-density distributions can be measured, respectively, using the time of flight technique and  atom noise in absorption images~\cite{theory_noise, expt_noise1,expt_noise2,expt_noise3,expt_noise4,expt_noise5}  or single atom resolved image~\cite{Jochim_expt2}.

	At finite low temperatures, the 2D QSF is expected to survive with quasi-long-range order and a Berezinskii-Kosterlitz-Thouless(BKT)-type transition can occur at a critical $T_{\rm BKT}$. In the strong coupling regime with deep quartet binding,
	we estimate $T_{\rm BKT}$  through the 2D quasi-condensation of bosons~\cite{Prokofev} as $T_{\rm BKT}/T_F^{Q}=\ln^{-1}[-\xi/(2\pi)\ln(\sqrt{4\pi}k_Fa)]$, with $\xi=380$ and $T_F^{Q}=k_F^2/[2(m_l+3m_h)]$. Here we have assumed the quartets interacting via a repulsive potential with range $\sim a$~\cite{Petrov3}. For $\ln(k_Fa)=-5\sim-2$, we obtain a slowing varying $T_{\rm BKT}/T_F^{Q}=0.18\sim0.26$. 
	 Given successful measurements of $T_{\rm BKT}$ in pairing superfluids of spin-$1/2$ Fermi gases~\cite{expt1,expt2,expt3,expt4}, we expect the BKT transition of QSF can also be explored in mass-imbalanced Fermi mixtures~\cite{K_Li1, K_Li2, K_Li3,Dy_K1, Dy_K2, Cr_Li, Cr_Li2, Cr_Li3}, for instance, by measuring the quartet momentum distribution similar to  Ref.~\cite{expt1}.
	
	In future, the present theory  could be further improved by incorporating the Pauli effect while treating Eq.~(\ref{wf_QSF}), which becomes more important as departing from strong coupling regime. 
Moreover, it is interesting to consider a general number ratio ($N_h/N_l\neq 3$), where QSF may coexist with other states and result in an even richer phase diagram. In addition, to be more relevant to ultracold experiments, it is desirable  to address the effects of finite $T$ and finite effective range over the whole interaction regime in a quasi-2D geometry. For a pure 3D system,  QSF is expected to appear at higher $\eta$ that can support a quartet in vacuum~\cite{Blume, Petrov}. In this case, the  true long range order of QSF can extend to finite $T$, and at deep bindings the critical $T_c$ approaches the  transition temperature for quartet condensation, i.e., $T_c/T_F^Q=0.44$.

	Finally, it is worth pointing out that the many-body phenomenon of QSF is deeply rooted in the highly non-trivial few-body physics, namely, the exact quartet formation of heavy-light fermions in vacuum. Based on this, we expect similar high-order superfluid to exist in a broad class of fermion systems with imbalanced (effective) masses. For instance, in a spin-orbit coupled atomic gas~\cite{review_soc} the lower helicity branch can possess a  large effective mass~\cite{Zheng_JPA}, which may serve as the heavy component to interact with other (light) fermion species. Another promising system is the monolayer transition metal dichalcogenides with charged excitons (trions)~\cite{trion_expt1,trion_expt2,trion_expt3}, where the mass-imbalanced mixture can consist of trions and electrons (or holes). Indeed, recent theories have revealed the existence of few-body clusters therein\cite{clusters_trion1,clusters_trion2}. 	For potentially exotic superfluids in these systems, our work suggests the few-to-many perspective as always a reliable route to approach them.

{\it Acknowledgements.}
	We are grateful to Yupeng Wang for insightful discussions. The work is supported by the National Natural Science Foundation of China (12074419, 12134015), and the Strategic Priority Research Program of Chinese Academy of Sciences (XDB33000000).

	\clearpage
	
	\onecolumngrid
	\vspace*{1cm}
	\begin{center}
		{\large\bfseries Supplemental Material}
	\end{center}
	\setcounter{figure}{0}
	\setcounter{equation}{0}
	\renewcommand{\figurename}{Fig.}
	\renewcommand{\thefigure}{S\arabic{figure}}
	\renewcommand{\theequation}{S\arabic{equation}}
	
	In this supplemental material, we provide more details on the derivation of density-density correlation function, the validity of our method in treating  quartet superfluid (QSF), and various phases competing with QSF in  the phase diagram (Fig.2(b) in the main text).
	
\section*{I.\ \ \ Density-density correlations of heavy fermions }
Under our treatment of QSF wavefunction (Eq.(3) in the main text), the mean density distribution of heavy fermions can be written as $\langle n_h({\cp k})\rangle\equiv \langle N_h({\cp k})\rangle/S$ with
\begin{eqnarray}
&&\langle N_h({\cp k})\rangle=\sum_{\langle\cp k' \cp k''\rangle}\frac{|\psi_{\cp k \cp k' \cp k''}|^2}{1+|\psi_{\cp k \cp k' \cp k''}|^2}. \label{Nf}
\end{eqnarray}
The two-body density distribution is $\langle n_h({\cp k}_0)n_h({\cp k})\rangle=\langle N_h({\cp k}_0)N_h({\cp k})\rangle/S^2$ with
\begin{eqnarray}
&&\langle N_h({\cp k}_0)N_h({\cp k})\rangle=\sum_{\cp k'}\frac{|\psi_{\cp k_0 \cp k \cp k'}|^2}{1+|\psi_{\cp k_0 \cp k \cp k'}|^2}+(\sum_{\langle\cp k' \cp k''\rangle}\frac{|\psi_{\cp k_0 \cp k' \cp k''}|^2}{1+|\psi_{\cp k_0 \cp k' \cp k''}|^2}-\sum_{\cp k'}\frac{|\psi_{\cp k_0 \cp k \cp k'}|^2}{1+|\psi_{\cp k_0 \cp k \cp k'}|^2})(\sum_{\langle\cp k' \cp k''\rangle}\frac{|\psi_{\cp k \cp k' \cp k''}|^2}{1+|\psi_{\cp k \cp k' \cp k''}|^2}-\sum_{\cp k'}\frac{|\psi_{\cp k_0 \cp k \cp k'}|^2}{1+|\psi_{\cp k_0 \cp k \cp k'}|^2}). \nonumber \\
\label{NN}
\end{eqnarray}
In above equation, the first term is contributed from the situation when ${\cp k}_0$ and ${\cp k}$ are from the same bracket in $|\Psi\rangle_{\rm QSF}$, i.e., within a quartet; and the second term is from when ${\cp k}_0$ and ${\cp k}$ are from different brackets, i.e., between the quartets. Introducing $A_{{\cp k}_0{\cp k}}$ as the intra-quartet correlation function:
\begin{equation}
A_{{\cp k}_0{\cp k}}=\sum_{\cp k'}\frac{|\psi_{\cp k_0 \cp k \cp k'}|^2}{1+|\psi_{\cp k_0 \cp k \cp k'}|^2}
\end{equation}
and in combination with Eq.~(\ref{Nf}), we can simplify Eq.~(\ref{NN}) as
\begin{eqnarray}
&&\langle N_h({\cp k}_0)N_h({\cp k})\rangle=\langle N_h({\cp k}_0)\rangle \langle N_h({\cp k})\rangle + A_{{\cp k}_0{\cp k}}(1-\langle N_h({\cp k_0})\rangle-\langle N_h({\cp k})\rangle)+A_{{\cp k}_0{\cp k}}^2.
	\end{eqnarray}
This gives the density-density correlation function (defined by Eq.8 in the main text) as
\begin{eqnarray}
&& D_h({\cp k}_0,{\cp k})=\frac{1}{S^2}\Big(A_{{\cp k}_0{\cp k}}(1-\langle N_h({\cp k_0})\rangle-\langle N_h({\cp k})\rangle)+A_{{\cp k}_0{\cp k}}^2\Big).
	\end{eqnarray}
Apparently, $D_h({\cp k}_0,{\cp k})$ is predominantly determined by the intra-quartet correlation  $l_{{\cp k}_0{\cp k}}$, and therefore they share similar structure in momentum space.

\section*{II.\ \ \ Validity for  the treatment of QSF}
	
	We have proposed a variational ansatz $|\Psi\rangle_{\rm QSF}$ (Eq.(3) in the main text) to describe QSF based on the idea of quartet (boson) condensation. However, it is quite complicated to exactly treat $|\Psi\rangle_{\rm QSF}$ given three independent momenta (${\cp k}_1$, ${\cp k}_2$, ${\cp k}_3$), or a momentum-triple, to label a quartet in each bracket. The presence of the ${\cp k}$-triples brings huge complexities in solving many-body problem mainly in view of two aspects:
	
	(i) the  annihilation of terms from multi-brackets due to Pauli effect, and
	
	(ii) the non-uniqueness of fermion occupation from multi-brackets.
	
	To simplify the problem, in this work we have neglected the contribution from above complexities (i,ii). We will evaluate the corrections from (i) and (ii) later.
	
	\begin{figure}[t]
		\includegraphics[width=10cm]{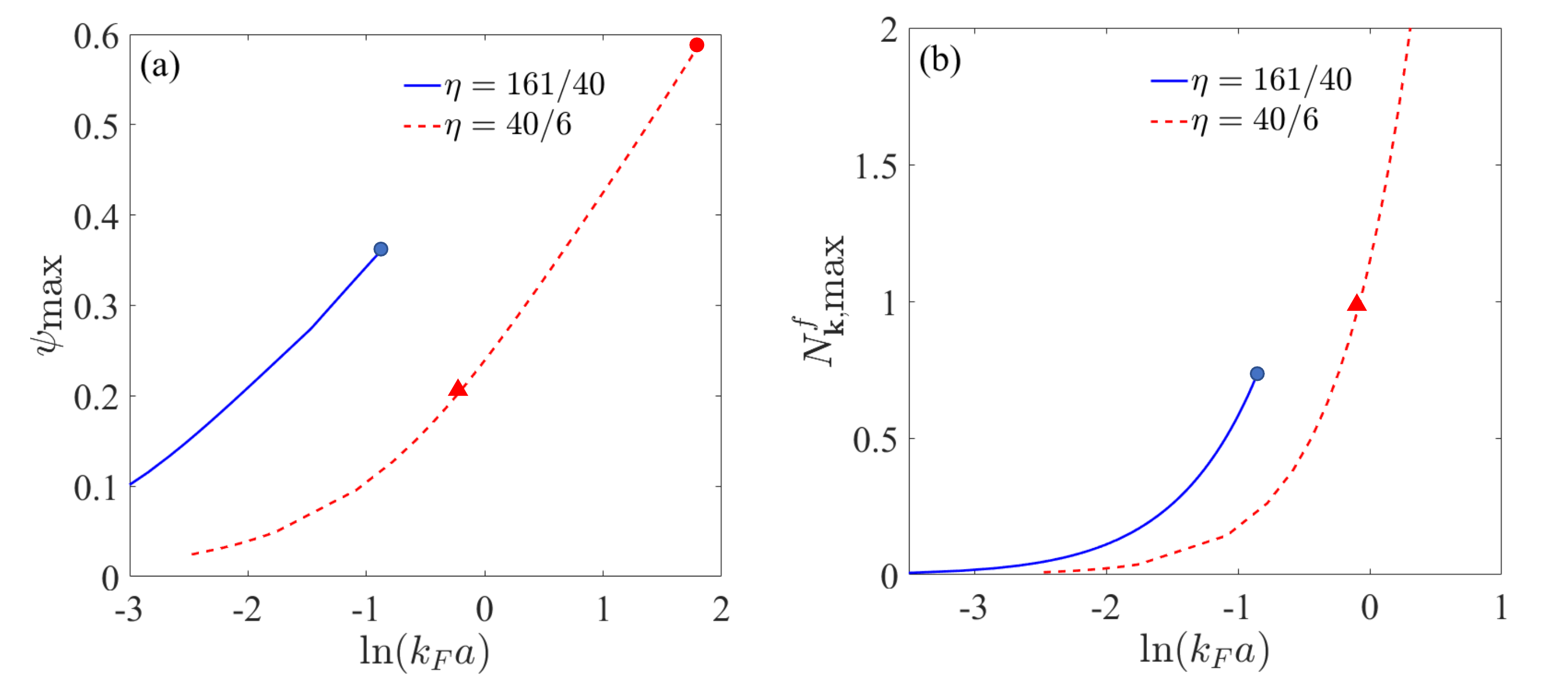}
		\caption{\label{fig_psi}(Color Online). (a)The largest $|\psi_{{\cp k}_1{\cp k}_2{\cp k}_3}|\equiv\psi_{\rm max}$ where $({\cp k}_1,\ {\cp k}_2,\ {\cp k}_3)$ form a regular triangle as shown in Fig.~3(a) of the main text. The solid and dashed lines are respectively for Dy-K and K-Li systems. (b) The largest $N^h_{\cp k}\equiv N^h_{\cp k,\textrm{max}}$, where ${\cp k}$ lies in one of the momentum-triple for $\psi_{\rm max}$. The triangular and circle points respectively denote the places where $N^h_{\cp k,\textrm{max}}$ starts to be larger than 1 and where a convergent solution terminates for QSF. }
	\end{figure}

	Under above treatment,  each bracket has its own normalization factor $1/\sqrt{1+\psi^2}$, and the ${\cp k}$-space distribution of heavy fermions can be written as
	\begin{equation}
		N^h_{\cp k}=\sum_{\langle {\cp k}_2 {\cp k}_3\rangle} \frac{|\psi_{{\cp k}{\cp k}_2{\cp k}_3}|^2}{1+|\psi_{{\cp k}{\cp k}_2{\cp k}_3}|^2}, \label{N_k}
	\end{equation}
	and the total number of heavy fermions is $N_h=3N_Q$ with
	\begin{equation}
		N_Q=\sum_{\langle{\cp k}_1{\cp k}_2{\cp k}_3\rangle} \frac{|\psi_{{\cp k}_1{\cp k}_2{\cp k}_3}|^2}{1+|\psi_{{\cp k}_1{\cp k}_2{\cp k}_3}|^2}. \label{N_Q}
	\end{equation}
	Here the notation $\langle \rangle$ means to avoid double counting of ${\cp k}$-pair in Eq.~(\ref{N_k}) or ${\cp k}$-triple in Eq.~(\ref{N_Q}). Note that in Eq.~(\ref{N_k}), the summation on two independent momenta (${\cp k}_2$ and ${\cp k}_3$) can easily result in a large $N^h_{\cp k}$ beyond unity, violating the Pauli principle of fermions. Such violation  is exactly because the neglect of Pauli effect in (i). 
	
	Now we roughly evaluate the correction from (i,ii) above, which can be seen easily from the 
	product of two brackets in $|\Psi\rangle_{\rm QSF}$:
	\begin{eqnarray}
		&&(1+\psi_{{\cp k}_1{\cp k}_2{\cp k}_3} l^{\dag}_{\cp Q} h^{\dag}_{{\cp k}_1}h^{\dag}_{{\cp k}_2}h^{\dag}_{{\cp k}_3})(1+\psi_{{\cp k}'_1{\cp k}'_2{\cp k}'_3} l^{\dag}_{\cp Q'} h^{\dag}_{{\cp k}'_1}h^{\dag}_{{\cp k}'_2}h^{\dag}_{{\cp k}'_3}) \nonumber\\
		&=& 1+\psi_{{\cp k}_1{\cp k}_2{\cp k}_3} l^{\dag}_{\cp Q} h^{\dag}_{{\cp k}_1}h^{\dag}_{{\cp k}_2}h^{\dag}_{{\cp k}_3} +\psi_{{\cp k}'_1{\cp k}'_2{\cp k}'_3} l^{\dag}_{\cp Q'} h^{\dag}_{{\cp k}'_1}h^{\dag}_{{\cp k}'_2}h^{\dag}_{{\cp k}'_3} + \psi_{{\cp k}_1{\cp k}_2{\cp k}_3} \psi_{{\cp k}'_1{\cp k}'_2{\cp k}'_3} l^{\dag}_{\cp Q} h^{\dag}_{{\cp k}_1}h^{\dag}_{{\cp k}_2}h^{\dag}_{{\cp k}_3} l^{\dag}_{\cp Q'} h^{\dag}_{{\cp k}'_1}h^{\dag}_{{\cp k}'_2}h^{\dag}_{{\cp k}'_3}, \label{product}
	\end{eqnarray}
	with ${\cp Q}=-{\cp k}_1-{\cp k}_2-{\cp k}_3$ and ${\cp Q}'=-{\cp k}'_1-{\cp k}'_2-{\cp k}'_3$. The complexities (i,ii) come from the last term of above equation.
	
	{\bf (i) The Pauli effect:}
	
	Take the heavy fermions for example, the Pauli principle will take effect when one of $\{{\cp k}_1,{\cp k}_2,{\cp k}_3\}$ is identical to one of $\{{\cp k}'_1,{\cp k}'_2,{\cp k}'_3\}$. For instance, when ${\cp k}_1={\cp k}'_1={\cp k}$, the last term of Eq.~(\ref{product}) will vanish. In this case and based on the two-bracket terms in Eq.~(\ref{product}), 
	the number distribution of heavy fermions at ${\cp k}$ is
	\begin{equation}
		N^h_{\cp k}= \frac{\psi_{{\cp k}{\cp k}_2{\cp k}_3}^2+\psi_{{\cp k}{\cp k}'_2{\cp k}'_3}^2}{1+\psi_{{\cp k}{\cp k}_2{\cp k}_3}^2+\psi_{{\cp k}{\cp k}'_2{\cp k}'_3}^2}.
	\end{equation}
	However, under the assumption that neglects the Pauli effect (see Eq.~(\ref{N_k})), we have
	\begin{equation}
		N^{h,(0)}_{{\cp k}}= \frac{\psi_{{\cp k}{\cp k}_2{\cp k}_3}^2}{1+\psi_{{\cp k}{\cp k}_2{\cp k}_3}^2} +  \frac{\psi_{{\cp k}{\cp k}'_2{\cp k}'_3}^2}{1+\psi_{{\cp k}{\cp k}'_2{\cp k}'_3}^2}.
	\end{equation}
	The discrepancy between 
	$N^h_{\cp k}$ and $N^{h,(0)}_{\cp k}$ directly reflects the correction brought by the Pauli effect, and obviously such discrepancy increases with $|\psi|$. By choosing $\psi_{{\cp k}{\cp k}_2{\cp k}_3}\sim \psi_{{\cp k}{\cp k}'_2{\cp k}'_3}\sim \psi_{\rm max}$ with $\psi_{\rm max}$ the maximum of $\psi$ in momentum space, now we  can estimate the largest (relative) correction to $N^h_{\cp k}$ as
	\begin{equation}
		\delta N^h_{\cp k} \equiv \frac{N^{h}_{\cp k}-N^{h,(0)}_{\cp k}}{N^{h,(0)}_{\cp k}} =-\frac{\psi_{\rm max}^2}{1+\psi_{\rm max}^2}. \label{correction}
	\end{equation}
	
	Eq.~(\ref{correction}) tells that the leading number correction by the Pauli effect is of the order $\sim \psi^2$. In Fig.~\ref{fig_psi}(a), we plot out the largest $\psi_{\rm max}$ as a function of $\ln(k_Fa)$ for both Dy-K and K-Li systems. One can see that $\psi_{\rm max}$ decays continuously as tuning the interaction to strong coupling limit ($\ln(k_Fa)\rightarrow -\infty$), where the Pauli effect is expected to take little role. To check this expectation, in Fig.~\ref{fig_psi}(b) we  show the largest $N^h_{\cp k}$ (based on Eq.~(\ref{N_k})) as functions of $\ln(k_Fa)$. One can see that as tuning $\ln(k_Fa)$ from strong to weak couplings, $N^h_{\cp k}$ gradually  increases, and for K-Li system with a larger mass imbalance,  $N^h_{\cp k}$ can be even over unity (triangular point), signifying the violation of Pauli principle. This can be attributed to the gradually increasing $\psi$ with $\ln(k_Fa)$, which leads to a larger relative correction to $N_{\cp k}^h$ according to Eq.~(\ref{correction}). However, as the system moving to strong coupling regime, both $\psi_{\rm max}$ and the largest $N^h_{\cp k}$ are well below unity,  ensuring the validity of neglecting Pauli effect in our treatment.
	
	{\bf (ii) The non-uniqueness of number occupation in $\cp k$-space:}
	
	It is noted that the last term in Eq.~(\ref{product}) can be produced from different brackets (or quartets). Specifically, it can be created from the product of $l^{\dag}_{\cp Q} h^{\dag}_{{\cp k}_1}h^{\dag}_{{\cp k}_2}h^{\dag}_{{\cp k}_3}$ and $l^{\dag}_{\cp Q'} h^{\dag}_{{\cp k}'_1}h^{\dag}_{{\cp k}'_2}h^{\dag}_{{\cp k}'_3}$, or the product of $l^{\dag}_{\cp Q} h^{\dag}_{{\cp k}_1}h^{\dag}_{{\cp k}'_2}h^{\dag}_{{\cp k}'_3}$ and $l^{\dag}_{\cp Q'} h^{\dag}_{{\cp k}'_1}h^{\dag}_{{\cp k}_2}h^{\dag}_{{\cp k}_3}$ as long as ${\cp k}_2+{\cp k}_3={\cp k}'_2+{\cp k}'_3$. This causes the non-uniqueness of multi-fermion occupation in momentum space. Such non-uniqueness will lead to the correlation between different brackets, and bring corrections to physical quantities. To evaluate such correction, we consider the term from four correlated brackets (here ${\cp k}_2+{\cp k}_3={\cp k}'_2+{\cp k}'_3$):
	\begin{eqnarray}
		&&(1+\psi_{{\cp k}_1{\cp k}_2{\cp k}_3} l^{\dag}_{\cp Q} h^{\dag}_{{\cp k}_1}h^{\dag}_{{\cp k}_2}h^{\dag}_{{\cp k}_3})(1+\psi_{{\cp k}'_1{\cp k}'_2{\cp k}'_3} l^{\dag}_{\cp Q'} h^{\dag}_{{\cp k}'_1}h^{\dag}_{{\cp k}'_2}h^{\dag}_{{\cp k}'_3})(1+\psi_{{\cp k}_1{\cp k}'_2{\cp k}'_3} l^{\dag}_{\cp Q} h^{\dag}_{{\cp k}_1}h^{\dag}_{{\cp k}'_2}h^{\dag}_{{\cp k}'_3})(1+\psi_{{\cp k}'_1{\cp k}_2{\cp k}_3} l^{\dag}_{\cp Q'} h^{\dag}_{{\cp k}'_1}h^{\dag}_{{\cp k}_2}h^{\dag}_{{\cp k}_3}) \nonumber\\
		&=& 1+\psi_{{\cp k}_1{\cp k}_2{\cp k}_3} l^{\dag}_{\cp Q} h^{\dag}_{{\cp k}_1}h^{\dag}_{{\cp k}_2}h^{\dag}_{{\cp k}_3} +\psi_{{\cp k}'_1{\cp k}'_2{\cp k}'_3} l^{\dag}_{\cp Q'} h^{\dag}_{{\cp k}'_1}h^{\dag}_{{\cp k}'_2}h^{\dag}_{{\cp k}'_3} +\psi_{{\cp k}_1{\cp k}'_2{\cp k}'_3} l^{\dag}_{\cp Q} h^{\dag}_{{\cp k}_1}h^{\dag}_{{\cp k}'_2}h^{\dag}_{{\cp k}'_3} +\psi_{{\cp k}'_1{\cp k}_2{\cp k}_3} l^{\dag}_{\cp Q'} h^{\dag}_{{\cp k}'_1}h^{\dag}_{{\cp k}_2}h^{\dag}_{{\cp k}_3} \nonumber\\
		&&+ (\psi_{{\cp k}_1{\cp k}_2{\cp k}_3} \psi_{{\cp k}'_1{\cp k}'_2{\cp k}'_3} +\psi_{{\cp k}_1{\cp k}'_2{\cp k}'_3} \psi_{{\cp k}'_1{\cp k}_2{\cp k}_3}) l^{\dag}_{\cp Q} h^{\dag}_{{\cp k}_1}h^{\dag}_{{\cp k}_2}h^{\dag}_{{\cp k}_3} l^{\dag}_{\cp Q'} h^{\dag}_{{\cp k}'_1}h^{\dag}_{{\cp k}'_2}h^{\dag}_{{\cp k}'_3}, \label{product4}
	\end{eqnarray}
	Note that the coefficient of the last term in above equation, $\psi_{{\cp k}_1{\cp k}_2{\cp k}_3} \psi_{{\cp k}'_1{\cp k}'_2{\cp k}'_3} +\psi_{{\cp k}_1{\cp k}'_2{\cp k}'_3} \psi_{{\cp k}'_1{\cp k}_2{\cp k}_3}$,  directly reflects the non-uniqueness of multi-fermion occupation and the correlation between different brackets (quartets). Since this term is of the order $\sim \psi^2$, its induced correction is also expected to be the order $\sim \psi^2$ at most. To check this, we evaluate the occupation number of a quartet staying at $\{{\cp Q}{\cp k}_1{\cp k}_2{\cp k}_3\}$:
	\begin{equation}
		N_{{\cp Q}{\cp k}_1{\cp k}_2{\cp k}_3}\equiv \langle l^{\dag}_{\cp Q} h^{\dag}_{{\cp k}_1}h^{\dag}_{{\cp k}_2}h^{\dag}_{{\cp k}_3}h_{{\cp k}_3}h_{{\cp k}_2}h_{{\cp k}_1}l_{\cp Q}\rangle,
	\end{equation}
	then based on Eq.~(\ref{product4}) we have
	\begin{equation}
		N_{{\cp Q}{\cp k}_1{\cp k}_2{\cp k}_3}=\frac{\psi_{{\cp k}_1{\cp k}_2{\cp k}_3}^2+(\psi_{{\cp k}_1{\cp k}_2{\cp k}_3} \psi_{{\cp k}'_1{\cp k}'_2{\cp k}'_3} +\psi_{{\cp k}_1{\cp k}'_2{\cp k}'_3} \psi_{{\cp k}'_1{\cp k}_2{\cp k}_3})^2}{1+ \psi_{{\cp k}_1{\cp k}_2{\cp k}_3}^2+\psi_{{\cp k}'_1{\cp k}'_2{\cp k}'_3}^2 +\psi_{{\cp k}_1{\cp k}'_2{\cp k}'_3}^2+\psi_{{\cp k}'_1{\cp k}_2{\cp k}_3}^2+
			(\psi_{{\cp k}_1{\cp k}_2{\cp k}_3} \psi_{{\cp k}'_1{\cp k}'_2{\cp k}'_3} +\psi_{{\cp k}_1{\cp k}'_2{\cp k}'_3} \psi_{{\cp k}'_1{\cp k}_2{\cp k}_3})^2}.
	\end{equation}
	However, under the original assumption we have
	\begin{equation}
		N^{(0)}_{{\cp Q}{\cp k}_1{\cp k}_2{\cp k}_3}=\frac{\psi_{{\cp k}_1{\cp k}_2{\cp k}_3}^2}{1+ \psi_{{\cp k}_1{\cp k}_2{\cp k}_3}^2}.
	\end{equation}
	Approximating all $\psi$ as $\psi_{\rm max}$, we obtain the largest (relative) correction as
	\begin{equation}
		\delta N_{{\cp Q}{\cp k}_1{\cp k}_2{\cp k}_3} \equiv \frac{N_{{\cp Q}{\cp k}_1{\cp k}_2{\cp k}_3}-N^{(0)}_{{\cp Q}{\cp k}_1{\cp k}_2{\cp k}_3}}{N^{(0)}_{{\cp Q}{\cp k}_1{\cp k}_2{\cp k}_3}} =\frac{\psi_{\rm max}^2}{(1+2\psi_{\rm max}^2)^2}. \label{correction2}
	\end{equation}
	
	We can see that the non-uniqueness from multi-brackets gives a positive correction to the number occupation (see Eq.~(\ref{correction2})), on contrary to the negative correction brought by the Pauli effect (Eq.~(\ref{correction})). However, their corrections are both of the order $\sim \psi^2$, which can be well controlled as long as $\psi\ll 1$ (especially valid in strong  coupling limit).

\section*{III.\ \ \ Competition between QSF and various other phases}
	
	Here we provide more details on the phases competing with QSF in the phase diagram (Fig.2(b) in the main text).

	\begin{figure}[h]
		\includegraphics[width=11cm]{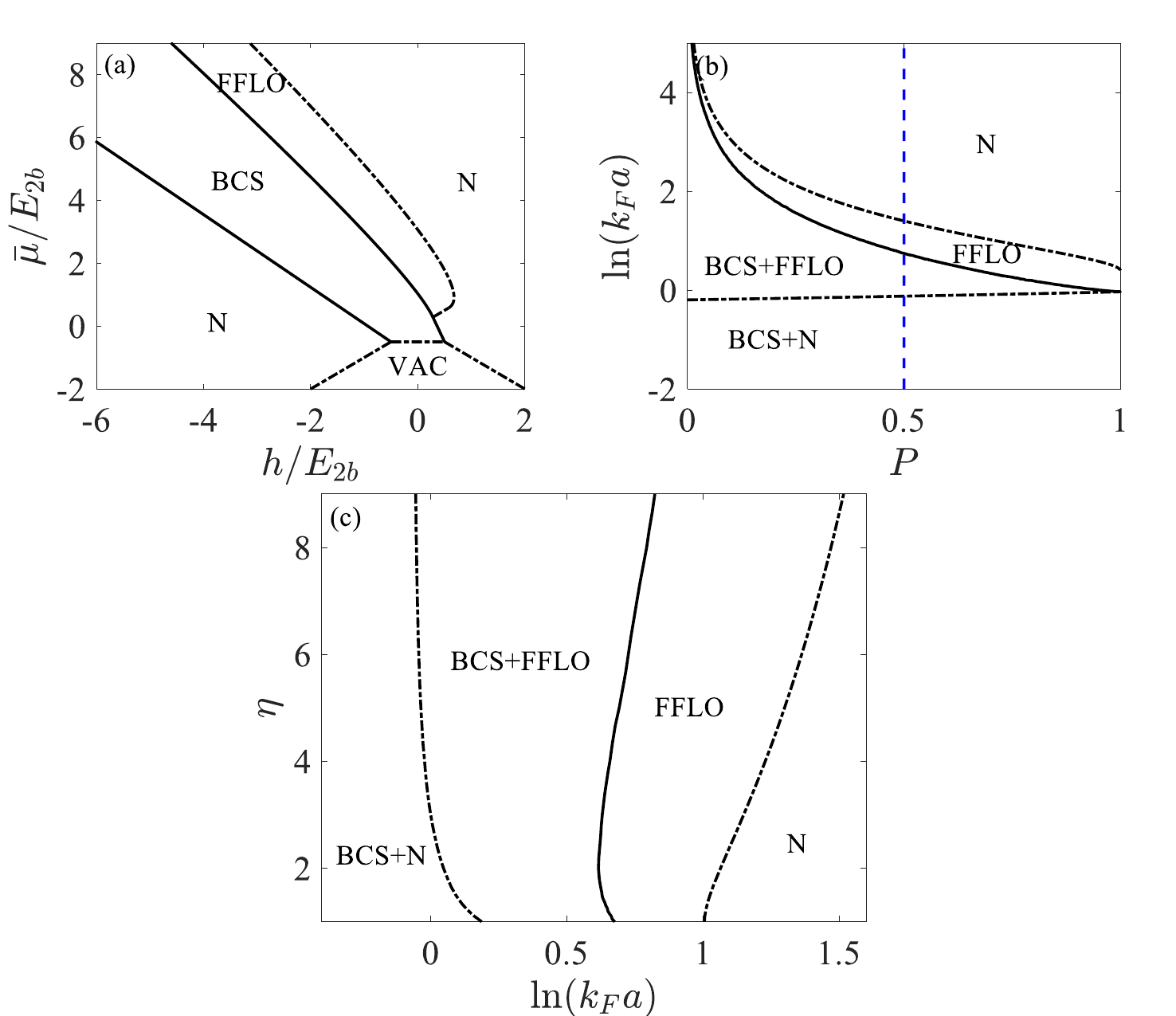}
		\caption{ (Color Online). Mean-field pairing phase diagrams in 2D. (a)Phase diagram in ($\bar{\mu}/E_{2b},h/E_{2b}$) plane for K-Li system. 'VAC', 'N', 'BCS' and 'FFLO' respectively correspond to the vacuum,   normal state, and  pairing superfluids with zero and finite pairing momenta. (b) Phase diagram in ($P, \ln(k_Fa)$) plane for K-Li system, with $P=(N_h-N_l)/(N_h+N_l)$ the polarization and $k_F$ is defined as the Fermi momentum of $(N_l+N_h)/4$ spinless fermions. 'BCS+N' (or 'BCS+FFLO') corresponds to the phase separation between BCS and normal (or between BCS and FFLO) states. The vertical line marks the number ratio $N_h/N_l=3$ (or $P=1/2$). (c)Phase diagram in ($\ln{k_Fa}, \ \eta$) plane with given number ratio $N_h/N_l=3$. In all diagrams, the solid (dash-dotted) line gives the first-order (continuous) phase boundary.}  \label{fig_pairing_diagram}
	\end{figure}
	
	\subsection*{A. Pairing superfluids}
	
	The pairing superfluids (such as BCS and FFLO) are described by the following wavefunction:
	\begin{equation}
		|\Psi\rangle_{\rm PSF} = \prod_{E_{{\cp k}+}<0} h^{\dag}_{{\cp k}+{\cp q}/2}  \prod_{E_{{\cp k}-}<0} l^{\dag}_{-{\cp k}+{\cp q}/2} \prod_{E_{{\cp k}\pm}>0} (1+\phi_{\cp k} h^{\dag}_{{\cp k}+{\cp q}/2} l^{\dag}_{-{\cp k}+{\cp q}/2} )
		|0\rangle, \label{wh_PSF}
	\end{equation}
	with $E_{{\cp k}\pm}$ the eigen-mode for pairing excitation and ${\cp q}$ the pairing momentum. In fact, Eq.~(\ref{wh_PSF}) is equivalent to the mean-field treatment of the Hamiltonian, which assumes a classical pairing field $\Delta\equiv (g/S) \sum_{\cp k}\langle h^{\dag}_{{\cp k}+{\cp q}/2} l^{\dag}_{-{\cp k}+{\cp q}/2} \rangle$. The BCS state corresponds to the ground state at ${\cp q}=0$ with equal species number $N_l=N_h$, while the FFLO state corresponds to the ground state at finite ${\cp q}$ with unequal species number $N_l\neq N_h$. Under the mean-field treatment, the gap and number equations can be written as
	\begin{eqnarray}
		\frac{S}{g}&=&-\frac{1}{2}\sum _{E_{{\cp k}\pm}>0}  \frac{1}{\sqrt{\left (\xi _{\cp k}^{+}-\bar{\mu} \right ) ^2+\Delta ^2}} \label{gap_eq2}\\
		N_l&=&\sum _{E_{{\cp k}-}<0 }1+\frac{1}{2}\sum _{E_{{\cp k}\pm}>0}[1-\frac{\xi _{\cp k}^{+}-\bar{\mu} }{\sqrt{\left (\xi _{k}^{+}-\bar{\mu} \right ) ^2+\Delta ^2} } ] \label{num_eq1}\\
		N_h&=&\sum _{E_{{\cp k}+}<0 }1+\frac{1}{2}\sum _{E_{{\cp k}\pm}>0}[1-\frac{\xi _{\cp k}^{+}-\bar{\mu} }{\sqrt{\left (\xi _{k}^{+}-\bar{\mu} \right ) ^2+\Delta ^2} } ] \label{num_eq2}
	\end{eqnarray}
	Where $\xi_\mathbf{k}^{\pm}=(\epsilon_{\mathbf{k+q/2}}^h \pm \epsilon_{\mathbf{-k+q/2}}^l)/2$,  $E_{{\cp k}\pm}=\pm (\xi_\mathbf{k}^{-} - h) + \sqrt{ (\xi _{k}^{+}-\bar{\mu} ) ^2+\Delta ^2}$, and the recombined chemical potentials are $\bar{\mu} =(\mu _h+\mu _l)/2$ and $h=(\mu _h-\mu _l)/2$.
	For the system of mass-imbalanced fermions in 2D, the mean-field phase diagrams have been obtained in Ref.~\cite{Conduit2}, where the FFLO state was extracted using the Ginzburg-Landau theory.

	
	In our work, we have taken the following steps to obtain the mean-field ground state with given number ratio $N_h/N_l=3$. As the first step, we fix the chemical potentials $\mu_l$ and $\mu_h$ and obtain the phase diagram in ($\bar{\mu},h$) plane by minimizing the thermodynamics potential $\Omega$ in terms of both $\Delta$ and ${\cp q}$. Note that the minimization in terms of $\Delta$ leads to the gap equation in Eq.~(\ref{gap_eq2}). The resulted phase diagram for K-Li system is shown in Fig.~\ref{fig_pairing_diagram}(a). The FFLO region therein deviates slightly from that in Ref.~\cite{Conduit2} due to different methods in extracting this phase.
	
	As the second step, we convert the ($\bar{\mu},h$) diagram into  diagram with fixed densities. In Fig.~\ref{fig_pairing_diagram}(b), we show the phase diagram in  ($P, \ln(k_Fa)$) plane for the same K-Li system, with $P=(N_h-N_l)/(N_h+N_l)$ the polarization and $k_F$ the Fermi momentum of $N_l$ light fermions. Here, the conversion is done by extracting the species densities at the phase boundaries of Fig.~\ref{fig_pairing_diagram}(a), and then mapping them into the phase boundaries in Fig.~\ref{fig_pairing_diagram}(b).
	For instance, the densities at the continuous FFLO-N boundary in Fig.~\ref{fig_pairing_diagram}(a)  give rise to the continuous FFLO-N boundary in Fig.~\ref{fig_pairing_diagram}(b), and the densities at the first-order BCS-FFLO boundary in Fig.~\ref{fig_pairing_diagram}(a)  give the upper and lower boundaries for the phase-separated BCS+FFLO state in Fig.~\ref{fig_pairing_diagram}(b).
	Given the ($P, \ln(k_Fa)$) diagram, we can easily identify the ground state for $N_h/N_l=3$ (or $P=1/2$), as shown by the vertical line in Fig.~\ref{fig_pairing_diagram}(b). One can see that as tuning $\ln(k_Fa)$ from weak to strong couplings, the system undergoes a sequence of transitions from a normal mixture to the phases involving various pairing superfluids. As the last step, by repeating the same procedure for different mass ratios, one can obtain the pairing phase diagram in ($\ln(k_Fa),\eta$) plane as shown in Fig.~\ref{fig_pairing_diagram}(c).

	\subsection*{B. Trimer liquid}
	
	A trimer liquid  of mass-imbalanced fermions has been proposed in 3D~\cite{Naidon2}, where the trimer refers to a bound state consisting of a light atom and two heavy fermions. Here we study a similar phase in 2D system, where a trimer in vacuum can emerge below dimer at mass ratio $\eta=3.34$~\cite{Pricoupenko2}. For the present case with $N_h/N_l=3$, we consider a homogeneous mixture of a trimer liquid  (composed by $N_l$ light fermions and $2N_l$ heavy ones) and a norma Fermi sea of rest heavy fermions (with number $N_h-2N_l=N_l$), as denoted by 'TL+N' in the phase diagram. For simplification, we further assume no interaction between trimer-trimer and between trimer-fermion. In this case, the total energy of this 'TL+N' state is given by $E_{\rm TL+N} = E_{\rm TL} +  E_{\rm N}$, where
	\begin{equation}
		E_{\rm TL}=N_l E_t(k_F) + \frac{N_l k_F^2}{8(2m_h+m_l)},  \ \ \ \ E_{\rm N}=\frac{N_l k_F^2}{4m_h}.  \label{trimerE}
	\end{equation}
	Here $E_t(k_F)$ is the energy of a trimer on top of the Fermi sea of $N_l$ heavy fermions (with Fermi momentum $k_F$), which can be obtained by solving the three-body equation subject to Pauli-blocking effect:
	\begin{equation}
		f_{\mathbf{k}_{2}}\left[\frac{S}{g}+\sum_{|\mathbf{k}_{1}|>k_F} \frac{1}{-E_t(k_F)+\epsilon_{-\mathbf{k}_{1}-\mathbf{k}_{2}}^{l}+\epsilon_{\mathbf{k}_{1} }^{h}+\epsilon_{\mathbf{k}_{2} }^{h}}\right]=\sum_{|\mathbf{k}_{1}|>k_F} \frac{f_{\mathbf{k}_{1}}}{-E_t(k_F)+\epsilon_{-\mathbf{k}_{1}-\mathbf{k}_{2}}^{l}+\epsilon_{\mathbf{k}_{1} }^{h}+\epsilon_{\mathbf{k}_{2} }^{h}}.
	\end{equation}
Due to the double degeneracy of trimer solutions in $p$-wave ($m=\pm 1$) channel, the trimers (of number $N_l=N_Q$) form two degenerate Fermi seas with Fermi momentum $k_F/\sqrt{2}$. This gives the total Fermi sea energy of trimers as the second term in $E_{\rm TL}$ (see Eq.~(\ref{trimerE})).
	
	To justify the existence of trimers in this system, we apply two conditions:  (1) $E_t(k_F)<0$ such that the trimer on top of the Fermi sea is well defined as a true bound state with localized wavefunction; and (2) $E_t(k_F)<E_d(k_F)+E_F^h$, where $E_d(k_F)$ is the lowest dimer energy on top of the heavy Fermi sea and $E_F^h=k_F^2/(2m_h)$ is the Fermi energy of heavy Fermi sea. The condition (2) is to make sure the trimers are stable (under collision) against the formation of dimers and free fermions on top of the heavy Fermi sea. To calculate $E_d(k_F)$, we have solved the two-body equation
	\begin{equation}
		\frac{S}{g}=-\sum _{|\cp k| > k_F} \frac{1}{\epsilon _{\cp K-\cp k}^{l}+\epsilon _{\cp k}^{h}-E_d(k_F) },
	\end{equation}
	and searched for the lowest $E_d(k_F)$ for all dimer momenta ${\cp K}$.
	

	\begin{figure}[h]
		\includegraphics[width=12cm]{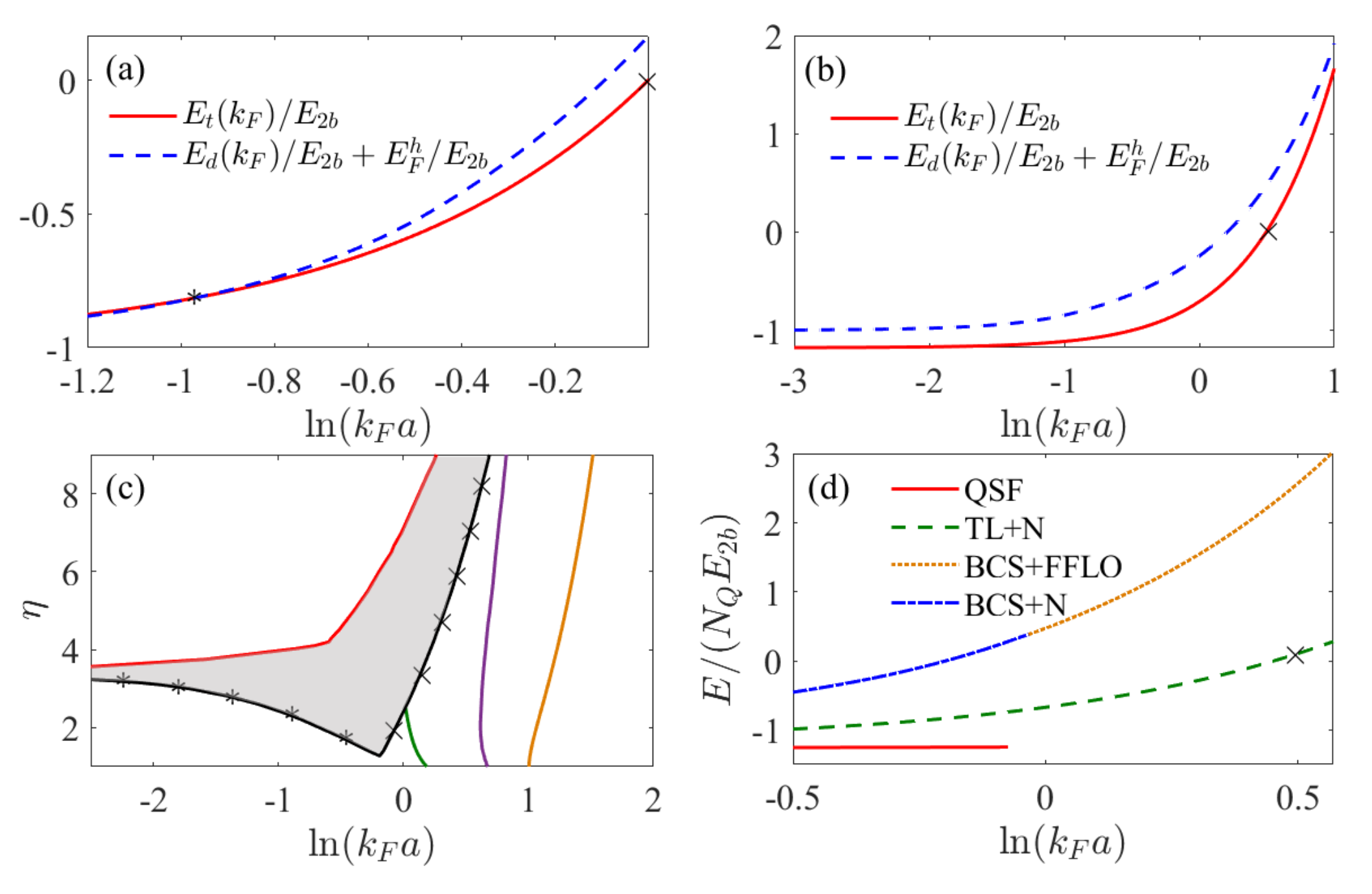}
		\caption{(Color Online). Mixture of a trimer liquid and a normal heavy Fermi sea ('TL+N'). (a) Trimer energy $E_t(k_F)$ (solid line) and the energy of lowest dimer-fermion state $E_d(k_F)+E_F^h$ (dashed line) as functions of $\ln(k_Fa)$ for mass ratio $\eta=2.4$. The star marks the location when $E_t(k_F)=E_d(k_F)+E_F^h$, and the crossing marks the location when $E_t(k_F)=0$. (b) The same as (a) except for mass ratio $\eta=40/6$.  (c) The region of  'TL+N' phase (gray area) in ($\ln(k_Fa),\eta$) plane. Its  boundaries are denoted by stars or crossings, following the same notations as in (a,b). 
			(d) Energy competition between QSF, TL+N and other phases involving pairing superfluids for mass ratio $\eta=40/6$.
		}  \label{fig_trimer_liquid}
	\end{figure}
	
	In Fig.~\ref{fig_trimer_liquid}(a,b), we show $E_t(k_F)$ (solid line) and $E_d(k_F)+E_F^h$ (dashed line) as functions of $\ln(k_Fa)$ for different mass ratios $\eta=2.4$ and $40/6$. It is found that even for small $\eta=2.4$ that cannot afford a trimer in vacuum, the trimer energy $E_t(k_F)$ can evolve below $E_d(k_F)+E_F^h$ as increasing $\ln(k_Fa)$ to some point, as marked by star in Fig.~\ref{fig_trimer_liquid}(a). This means that the Fermi-sea background will favor the trimer formation as compared to dimers. It is because the trimer involves two heavy fermions, which has a much larger scattering space than the dimer and thus is less affected by Pauli-blocking in the presence of heavy Fermi sea. The same effect can also be seen in Fermi polaron problem~\cite{Parish3,Parish4}, where the presence of heavy Fermi sea favors dressed trimers more than dressed dimers. For a larger $\eta=40/6$ that can well support a trimer (below dimer) in vacuum, as shown by Fig.~\ref{fig_trimer_liquid}(b), we can see that $E_t(k_F)$ is always lower than $E_d(k_F)+E_F^h$ for any coupling strength. However, $E_t(k_F)$ can cross zero as increasing $\ln(k_Fa)$ to some point, see crossings in Fig.~\ref{fig_trimer_liquid}(a,b), suggesting that the trimer is no longer a true bound state beyond this point.
	
	In Fig.~\ref{fig_trimer_liquid}(c), we have mapped out the region of 'TL+N' in ($\ln(k_Fa),\eta$) diagram satisfying both conditions (1) and (2), see the  gray area.  Moreover, we have checked that within this region the 'TL+N' state is indeed more energetically favored than all other available states, see Fig.~\ref{fig_trimer_liquid}(d) from the triangle to crossing points for a given mass ratio $\eta=40/6$. This justifies 'TL+N' as the unique ground state in this region.
	
	Finally, in above analysis we have neglected the repulsive interactions between trimer-trimer and trimer-fermion. The precise interaction strengths are still unknown up to date, which is expected to be quite small in strong coupling regime. The inclusion of these effects would increase the energy of  'TL+N' based on Eq.~(\ref{trimerE}) and thus  further shrink its region in the phase diagram.
	
	\subsection*{C. Pentamer liquid}
	
\begin{figure}[h]
		\includegraphics[width=7cm]{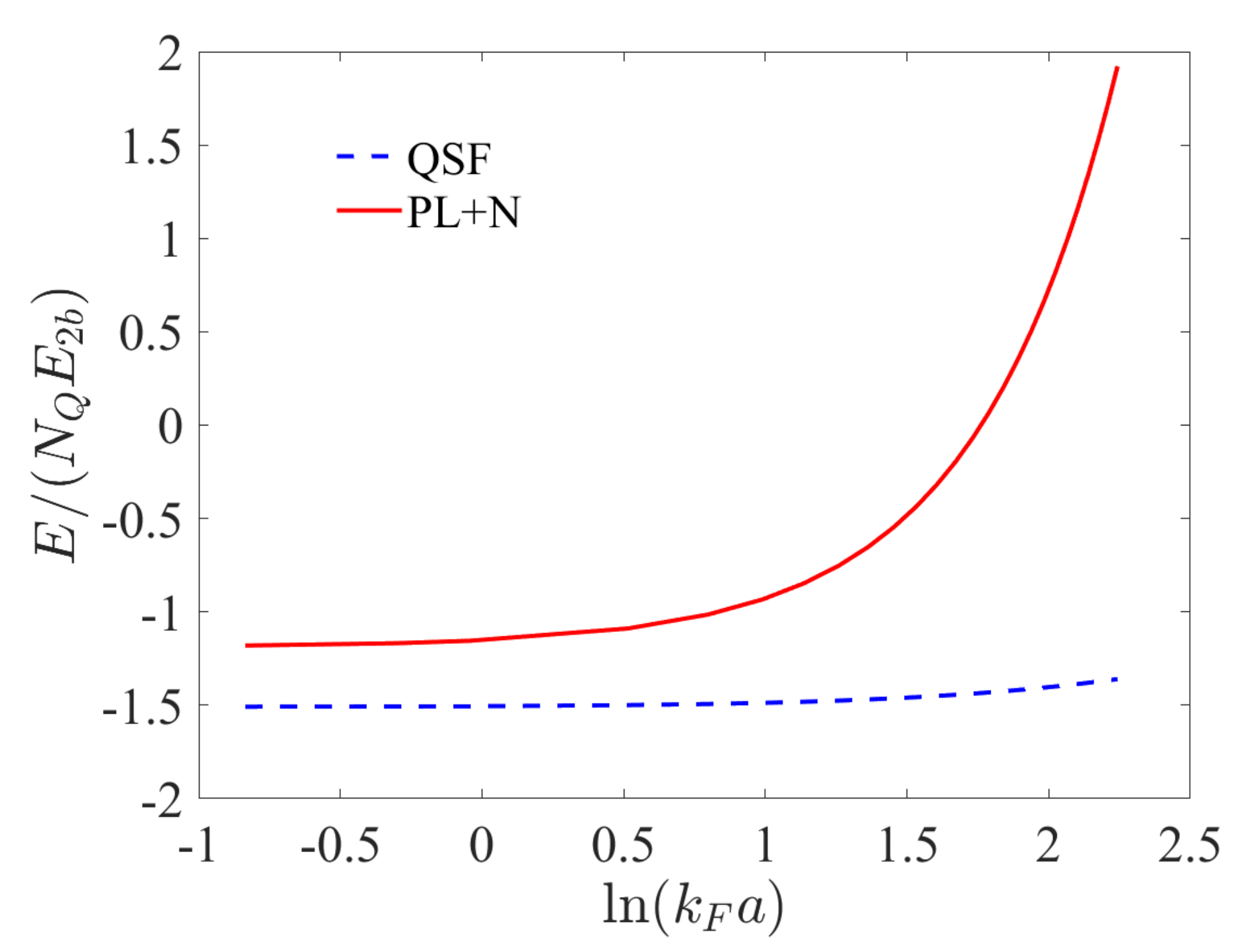}
		\caption{ (Color Online). Energy competition between the mixture of pentamer liquid and a normal light Fermi sea ('PL+N') and the quartet superfluid ('QSF') in strong coupling regime. In calculating $E_{\rm PL+N}$, we have use the vacuum $E_5^{(0)}$ in Ref.~\cite{Cui3} to replace $E_5$, which gives the lower bound to $E_{\rm PL+N}$. Here we consider the mass ratio $\eta=53/6$ with $E_5^{(0)}=-1.5826E_{2b}$ and $E_4^{(0)}=-1.5462E_{2b}$.  For the whole interaction regime, we can see QSF is always more energetically favorable than PL+N. }  \label{fig_pentamer}
	\end{figure}

	As a pentamer bound state (consisting of one light atom and four heavy fermions) can emerge in 2D vacuum at mass ratio $\eta=5.14$~\cite{Cui3}, this may give rise to a pentamer liquid in a many-body system. Here for a given number ratio $N_h/N_l=3$, 
	we consider a homogeneous mixture of a pentamer liquid (composed by $N_h$ heavy fermions and $N_h/4$ light fermions) and a norma Fermi sea of rest light fermions (with number $N_l-N_h/4=N_l/4$), denoted by 'PL+N' state. Its total energy  is given by $E_{\rm PL+N} = E_{\rm PL} +  E_{\rm N}$, where
	\begin{equation}
		E_{\rm PL}=(3N_l/4) E_5 + \frac{9}{16}\frac{N_l k_F^2}{4(4m_h+m_l)},  \ \ \ \ E_{\rm N}=\frac{1}{16}\frac{N_l k_F^2}{4m_l}.
	\end{equation}
	Here $E_5$ is the energy of a pentamer on top of the light Fermi sea, which should be higher than the energy of a pentamer in vacuum $E_5^{(0)}$. We have checked that even replace $E_5$ by $E_5^{(0)}$, $E_{\rm PL+N}$ cannot be lower than the energy of QSF for any mass ratio and coupling strength, see Fig.~\ref{fig_pentamer}. This can be attributed to the fact that the pentamer energy in vacuum is quite close to the quartet energy~\cite{Cui3}, i.e., $E_5^{(0)}\sim E_4^{(0)}$. Therefore, in strong coupling limit we have $E_{\rm PL+N}> (3N_l/4) E_5^{(0)} > E_{\rm QSF}\sim N_l E_4^{(0)}$, making 'PL+N' less favored than QSF in this regime. As the system moves to weak coupling regime, this phase is expected to be even less favored due to the Pauli-blocking effect ($E_5>E_5^{(0)}$).  As a result, such 'PL+N' cannot be a ground state in the phase diagram (Fig2.(b) in the main text).
	
	\subsection*{D. Effect of $S/a^2$ to the solution of quartet superfluid}
	
As we have neglected the Pauli effect in treating QSF ansatz, the summation on ${\cp k}$-triples in the number equation  brings a relevant quantity $S/a^2$ to the problem, which has been fixed  as $100$ in the main text. Here we show that the choice of $S/a^2$ will not change  QSF as ground state in strong coupling regime, but just  quantitatively affect its energy at intermediate couplings. Moreover, different choices of $S/a^2$ only bring slight modifications to the phase boundary of QSF in the phase diagram.

	\begin{figure}[h]
		\includegraphics[width=10cm]{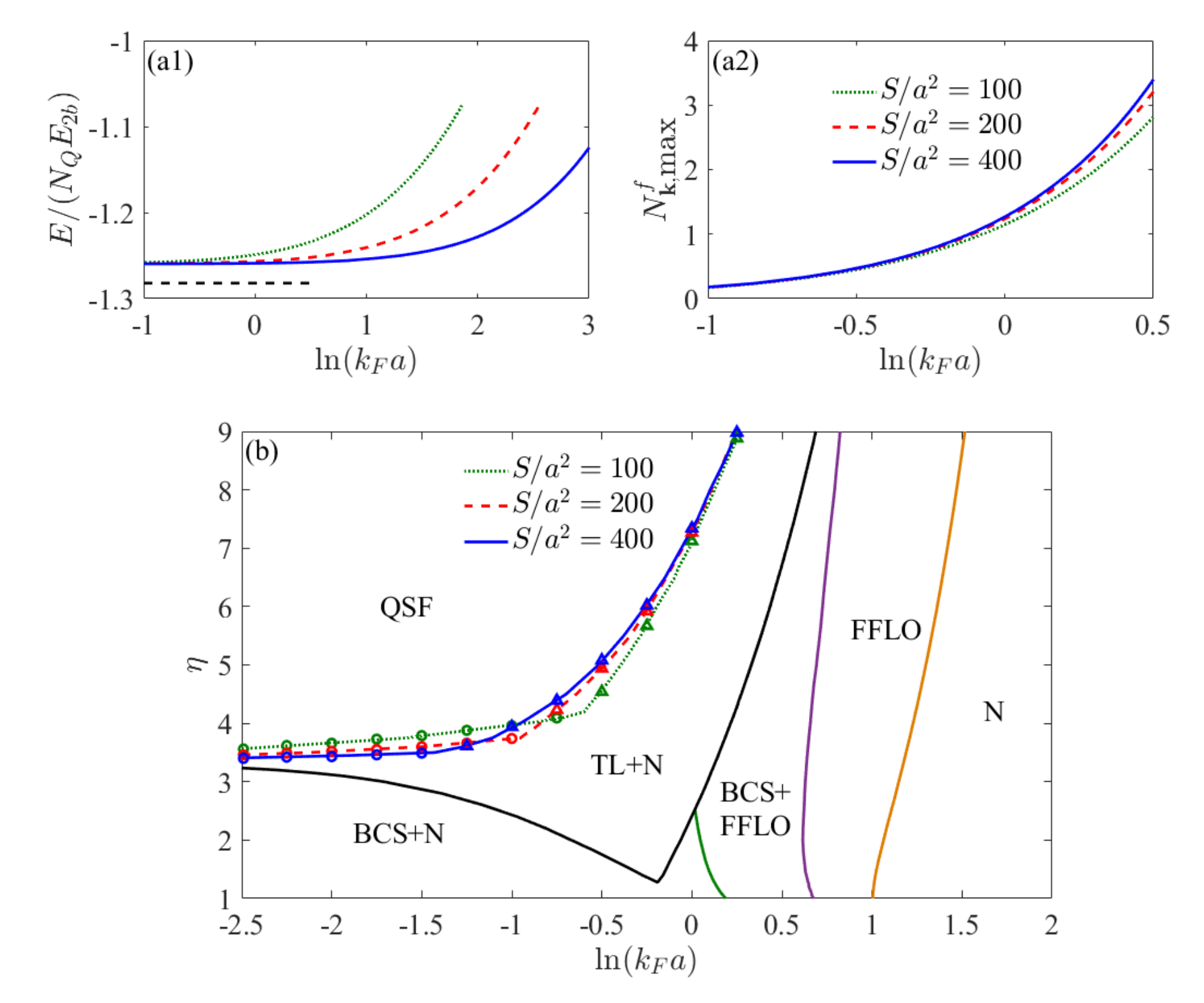}
		\caption{ (Color Online). Effect of the quantity $S/a^2$ to quartet superfluid (QSF). (a1) Energy per quartet $E/N_Q$(in unit of $E_{2b}$) and (a2) the largest $N^h_{\cp k}$ as functions of $\ln(k_Fa)$ for different $S/a^2=100,\ 200,\ 400$. Here we take the mass ratio $\eta=40/6$. The horizontal dashed line in (a1) denotes the quartet binding energy in vacuum.  (b) Phase boundaries of QSF for different $S/a^2=100,\ 200,\ 400$. In all plots, the circle (or triangle) point marks the location where the solution of QSF is no longer convergent (or no longer valid due to the violation of Pauli principle). }  \label{fig_S}
	\end{figure}

In Fig.~\ref{fig_S}(a1,a2), we take the K-Li mixture for example and plot out the energy per quartet $E/N_Q$ and  the largest $N^h_{\cp k}$ as functions of $\ln(k_Fa)$ for different $S/a^2=100,\ 200,\ 400$. One can see that different choices of $S/a^2$ will not alter the asymptotic behavior of  $E/N_Q\rightarrow E_Q$ (here $E_Q$ is the quartet binding energy in vacuum) in strong coupling limit. This guarantees QSF as the absolute ground state in this regime, which is more energetically favorable than all other competing phases. In addition, Fig.~\ref{fig_S}(a2) shows that the  violations of Pauli principle for different $S/a^2$ almost occur at the same $\ln(k_Fa)$. Therefore, the validity region of QSF is only slightly shifted by different $S/a^2$.

We have also checked the effect of $S/a^2$ for different $\eta$, and finally replotted the QSF phase boundary for different $S/a^2$, see Fig.~\ref{fig_S}(b). As expected, the QSF boundary is just slightly modified by $S/a^2$ ranging from $100$ to $400$. In particular,  the shift of phase boundary gets more invisible for higher mass imbalance, and we have confirmed this for even larger $S/a^2$. This justifies the yellow area in Fig.~2(b) of the main text as a trustable stability region for QSF as ground state.

\end{document}